\documentclass[twocolumn,showpacs,preprintnumbers,amsmath,amssymb]{revtex4}
\usepackage{graphicx}
\usepackage{dcolumn}
\usepackage{bm}
\input epsf

\begin{document}
\preprint{}
\title{Influence of magnetic fields on cold collisions of polar molecules}
\author{Christopher Ticknor and John L. Bohn}
\affiliation{JILA, University of Colorado, Boulder, CO 80309}
\date{\today}

\begin{abstract}
We consider cold collisions of OH molecules in the $^2\Pi_{3/2}$ grounds 
state, under the influnce of a magnetic field. We find that modest fields
of several thousand gauss can act to suppress inelastic collisions of 
weak-field seeking states by two orders of magnitude.  We attribute this 
suppression to two factors:  (i) An indirect coupling of the entrance and the 
exit channel, in contrast to the effect of an applied electric field; and 
(ii) the realtive motion of the entrance and exit scattering thresholds.
In view of these results, magnetic trapping of OH may prove feasible. 
\end{abstract}
\maketitle
\section{Introduction}

As the experimental reality of trapping ultra-cold polar molecules
approaches, a clear understanding is needed of how the molecules
interact in the trap environment.  On the most straightforward level,
collisions are essential for cooling the gas by either evaporative
or sympathetic cooling methods.  A high rate of elastic collisions
is desirable, while a low rate of exothermic, state-changing collisions
is essential if the cold gas is to survive at all.
Furthermore, a clear understanding of 2-body 
interactions allows one to construct a realistic model of the 
many body physics in this dilute system \cite{bec_rev}.  

One promising strategy for trapping ultracold molecules might
be to follow up on successes in trapping of cold atoms, and to
construct electrostatic \cite{Bethlem,Meerakker} or magneto static 
\cite{Weinstein} traps that can
hold molecules in a weak-field-seeking state.  
Cold collisions of polar molecules in this environment have
been analyzed in the past, finding that
inelastic collision rates were unacceptably high in the
presence of the electric field, limiting the possibilities
for stable trapping \cite{AA_PRA}.  Ref. \cite{AA_PRA} found that the
large inelastic rates were due to the strong dipole-dipole
interaction coupling between the molecules.  One important feature of the 
dipole-dipole interaction is its comparatively long range.
Even without knowing the details of the short-ranged molecule-molecule
interactions, the dipole forces alone were sufficient to change
the internal molecular states.  Indeed, a significant finding 
was that for weak-field seekers, the molecules are prevented from
approaching close to one another due to a set of long-range
avoided crossings.  Therefore, a reasonably accurate description
of molecular scattering may be made using the dipolar forces
alone \cite{AA_PRL}.

A complimentary set of theoretical analyses have considered the
problem of collisional stability of paramagnetic molecules
in a magnetostatic trap.  For example, the weak-field-seeking states
of molecules are expected to survive collisions with He buffer gas
atoms quite well \cite{BohnPRA2000,Krems1}.
Collisions of molecules with each other are also expected to preserve
their spin orientation fairly well, and hence
remain trapped \cite{AAPRA2001}.
However, this effect is mitigated in the presence of a magnetic field
\cite{Volpi,Krems2}.

So far, no one appears to have considered the influence of {\it magnetic}
fields on cold molecule-molecule collisions where both species
have {\it electric} dipole moments.
In this paper we approach this subject, by considering cold
OH($^2\Pi_{3/2}$)-OH($^2\Pi_{3/2}$) collisions in a magnetic field.
To the extent that the applied electric field is zero,
one might expect that dipole forces average to zero and thus do
not contribute to de-stabilizing the spin orientation.  It turns
out that this is not quite correct, and that dipole-dipole forces
still dominate long-range scattering.  However, applying a
suitably strong magnetic field turns out to mitigate this
effect significantly.
Interestingly, even in this case the residual second-order dipole
interactions are sufficiently strong to restrict scattering to 
large intermolecular separation 

\begin{figure}
\centerline{\epsfxsize=5.0cm\epsfysize=5.0cm\epsfbox{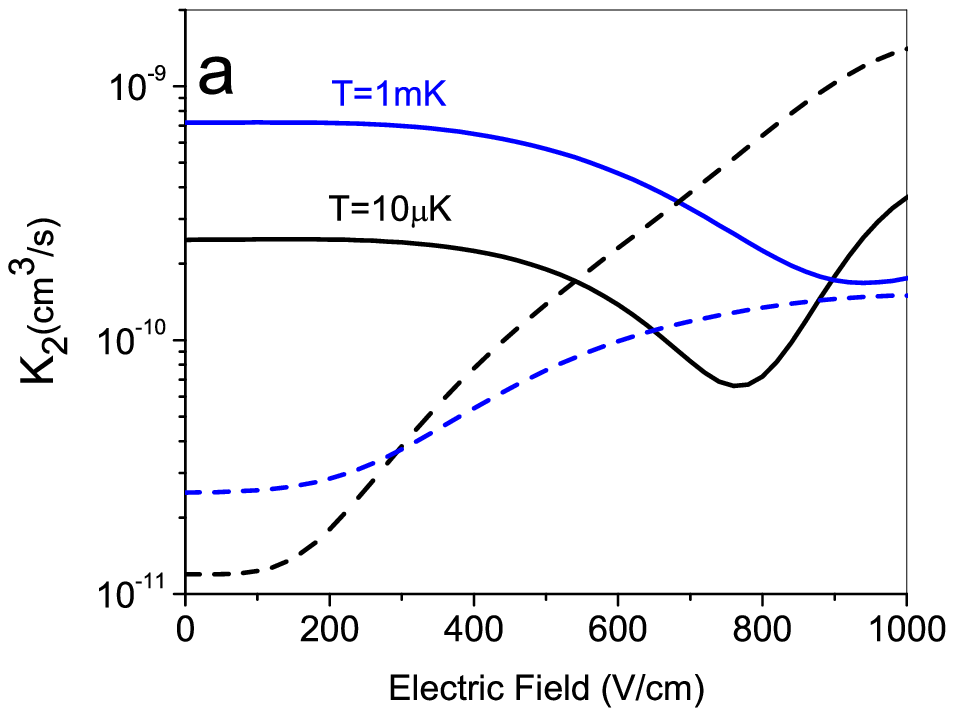}}
\centerline{\epsfxsize=5.0cm\epsfysize=5.0cm\epsfbox{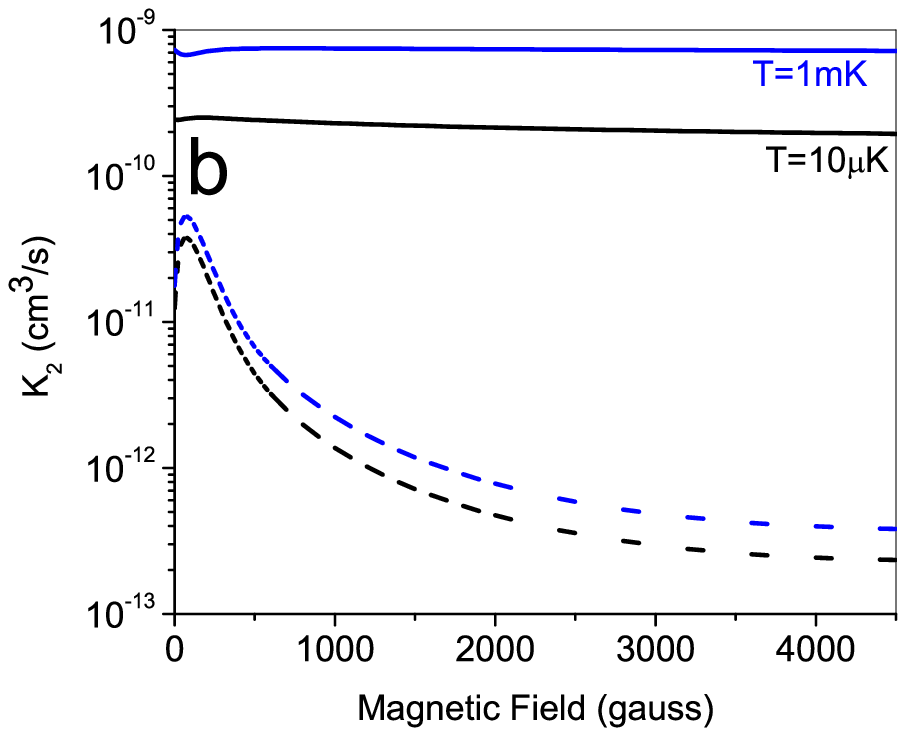}}
\caption{Thermally averaged rate constants for collisions of 
weak-field seeking sates of OH, as a fucntion of applied electric field
(a) and magnetic field (b).  In both cases, solid lines denote elastic 
scattering rates, while dashesd lines denote inelastic scattering rates.
Two temperatures are considered.  Applying an electric fields drives the
inelastic collisions rates up while applied magnetic field drives 
inelastic collision rates down.} 
\label{temp}
\end{figure}

The main result of the paper is summarized in Figure \ref{temp}, which
contrasts the influence of electric and magnetic fields.  Figure \ref{temp}(a)
plots the elastic (solid curves) and inelastic (dashed curves)
collision rate constants for OH molecules in their $|FM_F \epsilon\rangle
=|22-\rangle$, weak-field-seeking hyperfine state (for details on quantum 
numbes, see below).  As the electric field is increased, the elastic rate
constant grows to alarmingly large values, making the gas collisionally
unstable, as was shown in Ref. \cite{AA_PRA}.  Figure \ref{temp}(b), which 
is new to this paper, shows the analogous rate constants in a magnetic 
field (in both cases the field is assumed to lie along the
positive z axis of the laboratory reference frame).  In this case
the magnetic field has the effect of suppressing collisions,
all the way down to a rate constant of $2\times 10^{-13}$ cm$^3$/sec at 
fields
of $B=3000$ gauss.  These results are moreover fairly robust
against raising the temperature to the merely cold (not ultracold)
temperatures, $\sim1mK$, attainable in buffer-gas loading or Stark slowing
experiments.  This is good news for experiments -- it implies that
cooling strategies that rely on collisions may be feasible,
provided a suitably large bias magnetic field is applied.

Our main goal here is to analyze the suppression of rates in a
magnetic field.  The organization is as follows.
First we review the relevant molecular structure, and especially the Stark 
and Zeeman effects, to illustrate their complementary natures.
We then present an overview of the scattering model, including a 
review of the dipole-dipole interaction. 
Finally we present an analysis of the system in a magnetic 
field using a reduced channel model that encapsulates 
the essential collision physics.  Finally the model is qualitatively 
understood using the adiabatic representation.

\section{Electric versus magnetic fields applied to molecules}

Both the Stark and Zeeman effects in molecules have a similar form,
since both arise as the scalar product of a dipole moment with 
an external field.  Their influence on the molecule is quite
different, however, since they act on fundamentally different
degrees of freedom.  The electric field is concerned primarily
with where the charges {\it are} in the molecule, whereas the
magnetic field is concerned with where they are {\it going}.  
This is of paramount importance, since it implies that the electric
field is a true vector (odd under the parity operation), whereas
the magnetic field is a pseudovector (even under parity) \cite{Jackson}.
An electric field will therefore mix parity states of a molecule,
while a magnetic field will mix states only of a given parity.
This distinction is explored further in Ref. \cite{Freed}; here
we will only focus on those aspects of immediate relevance to our
project.

The rest of this paper will, fundamentally, restate this fact
in the context of scattering, and follow up the consequences
that arise from it.  To set the context of this discussion,
and to fix our notation, we first consider the molecules in 
the absence of external fields.

\subsection{Molecular structure in zero external field}

The OH radical has a complicated ground state structure which includes
rotation, parity, nuclear spin, electronic spin and orbital degrees 
of freedom.  We assume that the vibrational degrees
of freedom are frozen out at low temperatures, hence treat
the molecules as rigid rotors.  We further assume that
perturbations due to far away rotational levels are weak.
We do, however, include perturbatively the influence of the
$\Omega=1/2$ fine structure level, as described in Ref. \cite{AA_PRA}.

The electronic ground state of OH is $^2\Pi$,  with $\Omega=3/2$ .
OH is an almost pure Hund's case (a)
molecule, meaning the electronic degrees of
freedom are strongly coupled to the intermolecular axis. The
electronic state of the molecule in the $J$ basis is denoted by
$|JM_J\Omega\rangle|\Lambda\Sigma\rangle$ where $J$ is the total
electronic angular momentum, $M_J$ is its projection onto the lab
axis and $\Omega$ is $J$'s projection onto the molecular axis.
$\Sigma$ and $\Lambda$ are the projection of the electron's spin
and orbital angular momentum onto the molecular axis, and their
sum equals $\Omega$ ($=\Sigma+\Lambda$).  The electronic degrees
of freedom, $|\Lambda\Sigma\rangle$, will be suppressed for
notational simplicity because they are constant for all the collisional 
processes we consider. 

To describe the molecular wave function we assume a rigid rotor,
$\langle \alpha,\beta,\gamma |J M_J \Omega \rangle=\sqrt{2 J+1\over 8
\pi^2}D^{J\star}_{M_J\Omega}(\alpha,\beta,\gamma)$,
where$(\alpha,\beta,\gamma)$ are the Euler angles defining the molecular
axis and $D^{J\star}_{M_J\Omega}$ is a Wigner D-function.  
It is necessary for a $\Pi$ state molecule
to use the parity basis because OH has a 
$\Lambda$-doublet splitting which separates the two parity states ($e/f$). The
$\Lambda$-doublet arises from a coupling to a near by $\Sigma$ state. 
It is the coupling of the $\Sigma$ state to $\Pi$ state of the same 
parity which 
breaks the degeneracy of the two $\Pi$ parity states \cite{Brown}.
 
In the parity basis the molecular wave function is written 
\begin {equation}
\label{Jparitybasis}
|J M_J\bar
\Omega\pm \rangle=\left({|J M_J\Omega \rangle+\epsilon|J M_J
-\Omega \rangle\over \sqrt{2}}\right).
\end{equation}
 Where $\epsilon=+(-)$
represents the $e$ ($f$) state, and $\bar\Omega=|\Omega|$.  It 
should be noted that the sign of $\epsilon$ is not the parity, rather 
parity is equal to $\epsilon(-1)^{(J-1/2)}$ \cite{Brown}.  Thus for 
the ground state of OH 
where $J=3/2$, parity is equal to $-\epsilon$.  Throughout the paper we use 
$\pm$ denote the sign of $\epsilon$, not parity.  
In the parity basis there is no dipole moment, because this basis is a linear
combination of electric dipole ``up'' and ``down''.  This fact has important 
implications for the dipole-dipole interaction.

Including the hyperfine structure is important because the most dominant
loss processes are those that change the hyperfine quantum number of 
one or both of the scattering molecules.  
The hyperfine structure arises from interaction of the electronic
spin with the nuclear spin ($I$) which must then be included
in the molecular basis set.  In the hyperfine basis the OH states
are $|FM_F\pm\rangle$, where ${\bf F}={\bf J+I}$ and $M_F$ is its projection
onto the lab axis. Here we suppress $\bar \Omega$ in the notation,
as its value is understood.
To construct basis functions with quantum number $F$
we expand in Clebsh-Gordon coefficients:
\begin{equation}
\label{Fparitybasis}
|FM_F\pm\rangle=\sum_{M_J M_I} |J M_J\bar
\Omega\pm \rangle |IM_I\rangle\langle J M_J IM_I
|FM_F\rangle.
\end{equation}
 
Relevant molecular energy scales for the scattering problem are the  
$\Lambda$-doublet splitting which is $\Delta \sim 0.0797 K$, the hyperfine
splitting is $\Delta_{hf}\sim0.0038 K$. OH also has an electric dipole moment
is $\mu\sim1.668 D$.  Throughout this paper we use Kelvin 
$(K)$ as the Energy unit, except in the instances of thermally averaged
observables.  For reference, $0.64K=1 cm^{-1}$.

\subsection{Stark Effect in OH}

As noted above, the distinguishing feature of the Stark effect is
that it mixes molecular states of opposite parity separated by the 
$\Lambda$ doublet splitting.  A consequence of
this is that the Stark energies vary quadratically with electric field
at low fields, and linearly only at higher fields.  The field where
this transition occurs is given roughly by equating the field's
effect ${\vec \mu_{E}}\cdot {\vec E}$ to the $\Lambda$ doublet splitting
(here ${\vec \mu_{E}}$ is the molecule's electric dipole moment, and 
${\vec E}$ is the field.  In OH, this field is approximately
$\emph{E}_0\sim\Delta/2\mu_\emph{E}\sim1000(V/cm)$.)

The Stark Hamiltonian has the form
\begin{equation}
H_{S}=-{\vec\mu_{{\emph E}}}\cdot {\emph\bf E} \label{Stark}
\end{equation}
where we take the field to be in the $\hat z$ direction.  
In the basis in which $\Omega$ has a definite sign, 
the matrix elements are well known \cite{Townes}:
\begin{equation}
\label{StarkJbasis}
\langle JM_J \Omega|H_{S} |J M_J\Omega\rangle={-\mu_\emph{E} 
\emph{E} \Omega M_J\over
J(J+1)}.
\end{equation}
In the Stark effect there is a degeneracy between states with 
the same sign of $\Omega M_J$, meaning $\pm M_J$ are degenerate in 
an electric field.  We can recast the Stark 
Hamiltonian into the $J$-parity basis set from Eqn. (\ref{Jparitybasis}).  
Doing so, we find
\begin{equation}
\langle J M_J \bar\Omega\epsilon|H_{S} |J M_J
\bar\Omega\epsilon^\prime\rangle={-\mu_\emph{E}\emph{E} \bar\Omega M_J\over
J(J+1)}\left({1-\epsilon\epsilon^\prime\over 2}\right).\label{starkjp}
\end{equation}
In this expression,  the factor $(1-\epsilon \epsilon^\prime)/2$  
explicitly represents the electric field coupling between states of 
opposite parity, since it vanishes for $\epsilon = \epsilon^{\prime}$.

\begin{figure}
\centerline{\epsfxsize=5.0cm\epsfysize=5.0cm\epsfbox{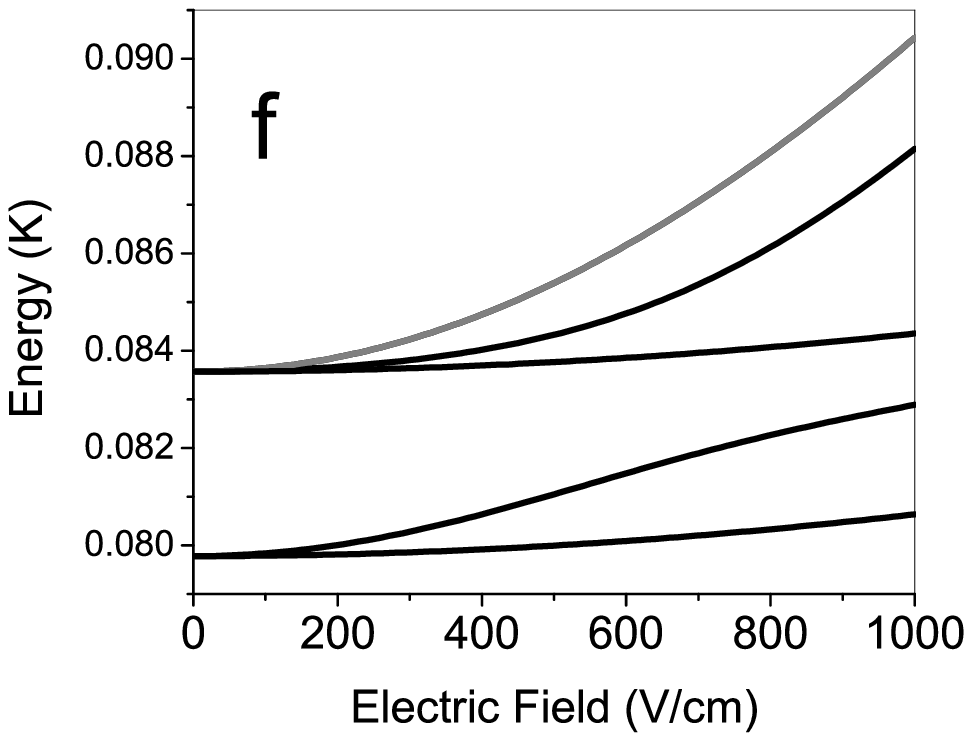}}
\centerline{\epsfxsize=5.0cm\epsfysize=5.0cm\epsfbox{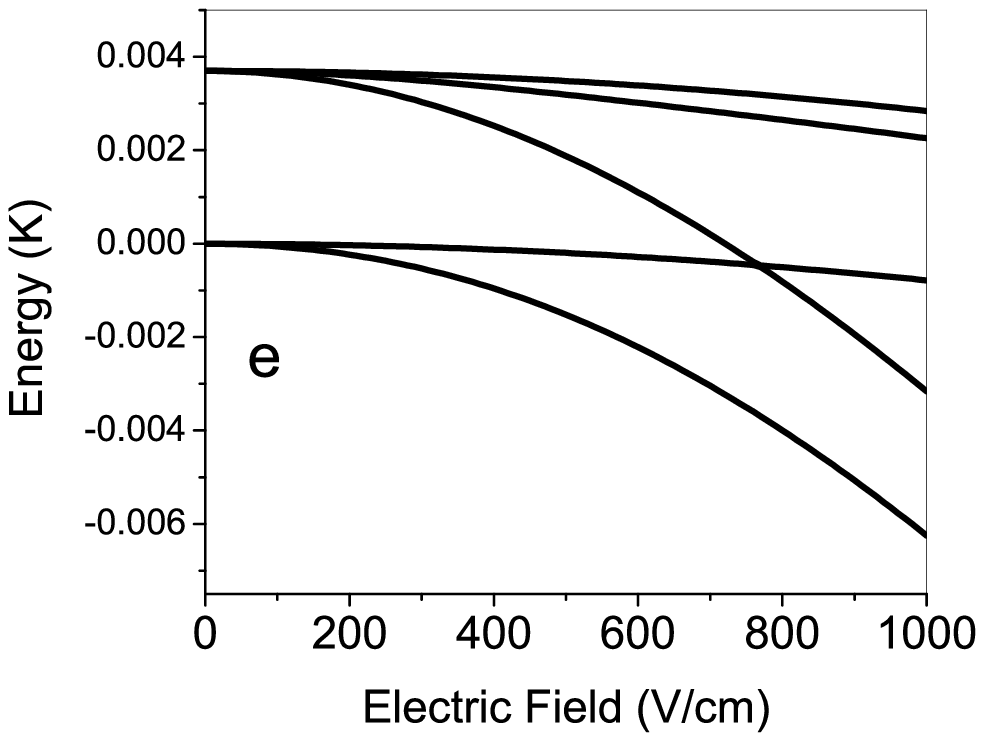}}
\caption{Stark effect for the ground state of OH
with the hyperfine structure accounted for.  In zero field 
the f states and the e states
are separated by the $\Lambda$-doublet energy.  The Gray line indicates the 
state of interest for our analysis, the $|22-\rangle$ state.  An important
feature of this interaction is that the opposite parity states repel and 
thus like parity state stay close together in energy.
} \label{starkenergy}
\end{figure}

Finally, using the definition of the $F$-parity basis in Eqn. 
(\ref{Fparitybasis}), we arrive at the working matrix elements of
the Stark effect:
\begin{eqnarray}
\langle F M_F \epsilon|H_{S} |F^\prime M_F\epsilon^\prime\rangle=
-{\mu_\emph{E} \emph{E}}\left({1+\epsilon\epsilon^\prime
(-1)^{J+J^\prime+2\bar\Omega+1}\over 2}\right)\nonumber\\
\times(-1)^{J+J^\prime+F+F^\prime+I-M_F-\bar\Omega+1}
[F,F^\prime,J,J^\prime]\nonumber\\
\times\left(\begin{array}{ccc}
J&1&J^\prime\\
-\bar\Omega&0&\bar\Omega^\prime
\end{array}\right)
\left(\begin{array}{ccc}
F^\prime&1&F\\
M_F&0&-M_F
\end{array}\right)
\left\{
\begin{array}{ccc}
F&F^\prime&1\\
J^\prime&J&I
\end{array}
\right\}.\label{starkhf}
\end{eqnarray}
In this notation $[j_1,j_2,...]=\sqrt{(2j_1+1)(2j_2+1)(...)}$. 
Figure \ref{starkenergy} shows the energy levels of 
OH in the presence of an electric field. Both parity states are shown, labeled 
$e$ and $f$.  An essential point of Fig. \ref{starkenergy} is that  
the $e$ and $f$ states repel as the electric field in increased.  This means 
the all of the $f$ ($e$) states increase (decrease) in energy as the field in
increased, implying that states of the same parity stay close
together in energy as the field is increased.  This fact has 
a crucial effect on the inelastic scattering as we will show.

The highest-energy state in Fig. \ref{starkenergy} is the stretched
state with quantum numbers $|FM_F \epsilon \rangle  =$
$|2 2 - \rangle$.  It is this state
whose cold collisions we are most interested in, because i) it is
weak-field seeking, and ii) its collisions at low temperature 
result almost entirely from long-range dipole-dipole interactions
\cite{AA_PRA}.  Molecules in this state will suffer inelastic 
collisions to all of the other internal states shown.  
The rate constant shown in Figure \ref{temp} is the sum of
all rate constants for all such processes.

\subsection{Zeeman Effect in OH}

When OH is in an external magnetic field the electron's
orbital motion and intrinsic magnetic dipole moment both
interact with the field.  The interaction is described by the Zeeman 
Hamiltonian which is
\begin{equation}
H_{Z}=-{\bf\mu}_B\cdot{\bf B}=\mu_0({\bf L}+g_e{\bf S})\cdot{\bf B}.
\end{equation}
Here $\mu_0$ is the Bohr magneton and $g_e$ is the electron's g factor
($g_e\sim 2.002$). As above, we assume the field to be in the laboratory 
$\hat z$ direction.  In the $J$ basis, the
Zeeman Hamiltonian takes the form \cite{herzberg}
\begin{equation}
\langle J M_J \Omega|H_{Z}| J M_J \Omega \rangle
={\mu_0 B (\Lambda+g_e \Sigma)\Omega M_J \over J(J+1)}.
\end{equation}
This is quite similar to the equivalent expression (\ref{StarkJbasis})
for the Stark effect, except that the electron's $g$-factor plays 
a role.  Interestingly, for a $^2\Pi$ state the prefactor
$(\Lambda+g_e \Sigma)\Omega$ is always greater than zero.
 We now recast the Zeeman interaction into the 
$J$-parity basis set (\ref{Jparitybasis}).  This gives us
\begin{equation}
\label{ZeemanJparitybasis}
\langle J M_J \bar\Omega\epsilon|H_{Z}| J M_J \bar\Omega\epsilon^\prime\rangle
={\mu_0 B \left(\bar\Lambda+g_e \bar\Sigma\right)\bar\Omega M_J \over J(J+1)}
\delta_{\epsilon\epsilon^\prime}
\end{equation}
for $\bar\Omega = 3/2$ states, and
\begin{equation}
\langle J M_J \bar\Omega\epsilon|H_{Z}| J M_J \bar\Omega\epsilon^\prime\rangle
={\mu_0 B \left(\bar\Lambda-g_e \bar\Sigma\right)\bar\Omega
M_J \over J(J+1)}
\delta_{\epsilon\epsilon^\prime}
\end{equation}
for $\bar\Omega=1/2$ states.
Notice that for $\bar\Omega=1/2$, the orbital and spin contributions
to the molecular magnetic moment nearly cancel, to within the 
deviation of $g_e/2$ from one.  For the $\bar\Omega=3/2$ states of interest
to us, however, the magnetic moment remains large. 

\begin{figure}
\centerline{\epsfxsize=4.0cm\epsfysize=4.0cm\epsfbox{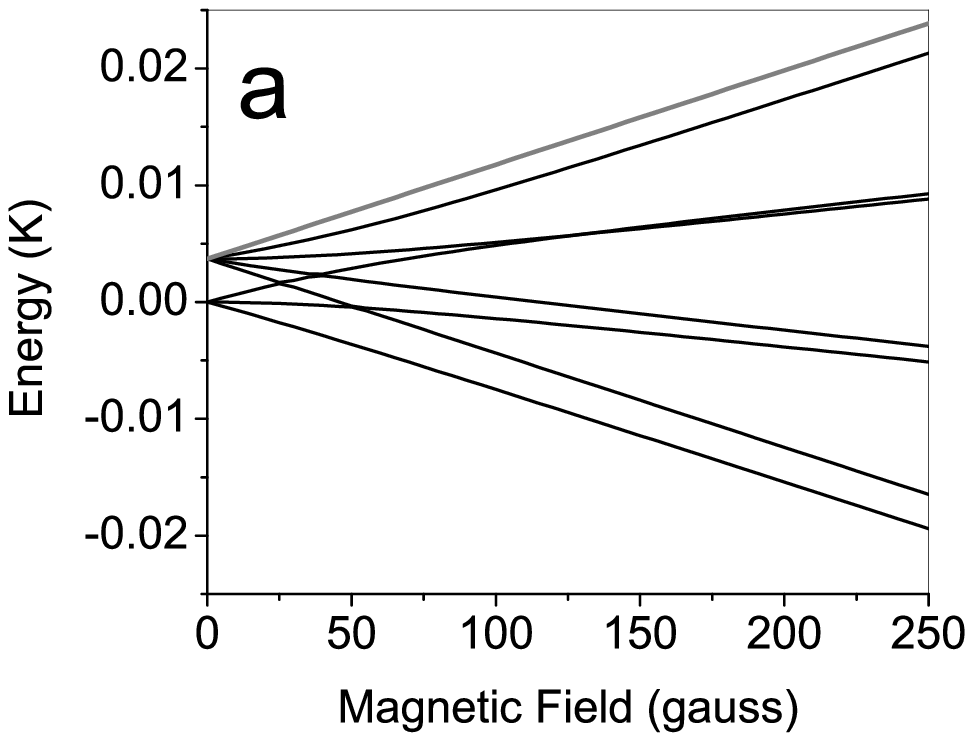},
\epsfxsize=4.0cm\epsfysize=4.0cm\epsfbox{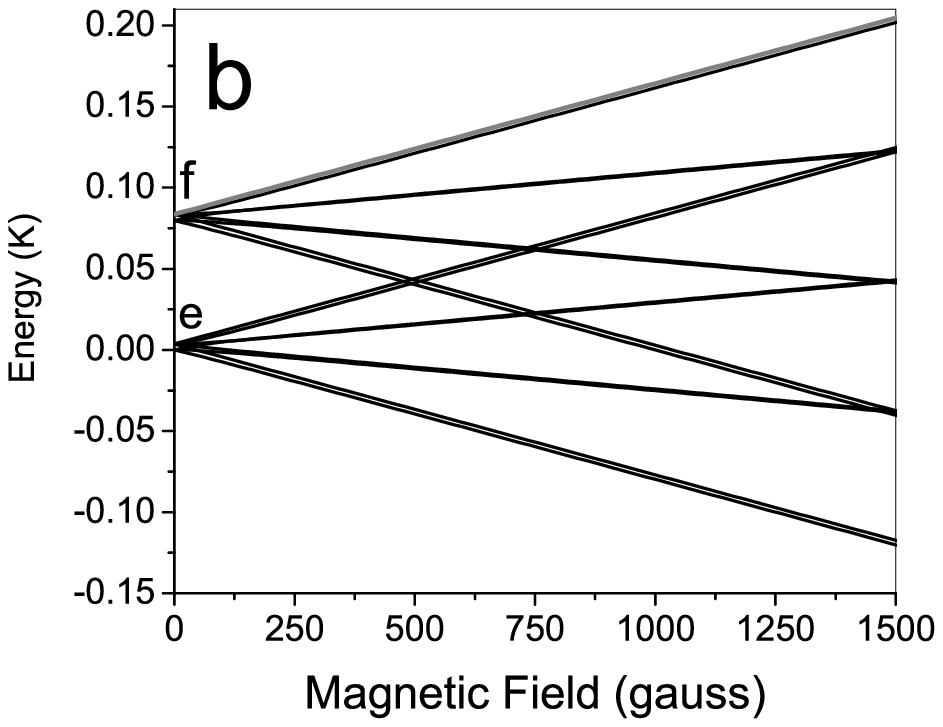}}
\caption{Zeeman effect for the ground state of OH, in low (a) and
high (b) fields.  This plot is the same for
both the $e$ and $f$ states for zero electric field because the Zeeman 
interaction respects preserves parity.  The $|22-\rangle$
state is indicated in gray.} \label{zeeman}
\end{figure}

The key feature of the Zeeman matrix element (\ref{ZeemanJparitybasis}) is
that it is diagonal in $\epsilon$, in contrast to the Stark matrix
element.  This trait persists in the hyperfine basis as well,
where the matrix elements are
\begin{eqnarray}
&&\langle F M_F \epsilon|H_{Z} |F^\prime M_F\epsilon^\prime\rangle=
\nonumber\\
&&{\mu_0B(\bar\Lambda+g_e\bar\Sigma)}
\left({1
+\epsilon\epsilon^\prime(-1)^{(J+J^\prime+2\bar\Omega)}\over 2}\right)
\nonumber\\
&&\times(-1)^{J+J^\prime+F+F^\prime+I-M_F-\bar\Omega+1}
[F,F^\prime,J,J^\prime]
\\ \nonumber
&&\times
\left(\begin{array}{ccc}
J&1&J^\prime\\
-\bar\Omega&0&\bar\Omega^\prime
\end{array}\right)
\left(\begin{array}{ccc}
F^\prime&1&F\\
M_F&0&-M_F
\end{array}\right)
\left\{
\begin{array}{ccc}
F&F^\prime&1\\
J^\prime&J&I
\end{array}
\right\}.\label{zhf}
\end{eqnarray}
Figure \ref{zeeman} shows the Zeeman energies in the hyperfine basis,
for low (Fig. \ref{zeeman}(a)) and high (Fig. \ref{zeeman}(b)) fields.
For OH in the $^2\Pi_{3/2}$ state, the parity factor 
$\left( {1+\epsilon\epsilon^\prime(-1)^{(J+J^\prime+2\bar\Omega)}
\over 2}\right)$ reduces simply to $\delta_{\epsilon\epsilon^\prime}$.
Because the magnetic field respects parity, figure \ref{zeeman} (b) 
amounts to
two copies of the same energy level diagram, separated in energy
by the lambda doublet energy.
For small magnetic fields the molecular $g$-factor is 
$g_{mag}^{OH}\propto(F^2+J^2-I^2)$, and is always positive for OH.
This is in contrast to the low field magnetic moment of alkali atoms which is
$g_{mag}^{alkali}\propto(F^2-J^2-I^2)$ (and where $J$, of course,
refers to the sum of orbital and spin angular momenta).
In Eqn. (\ref{zhf}) for $\bar\Omega=1/2$ the factor 
$\bar\Lambda+g_e\bar\Sigma$ goes to $\bar\Lambda-g_e\bar\Sigma$.  

\section{Scattering Hamiltonian}

A complete potential energy surface for the interaction of
two OH molecules, including the relatively long-range
part most relevant to cold collisions, is at present unavailable.
Certain aspects of this surface have, however, been discussed 
\cite{surface}.
For the time being, we will follow our previous approach of
focusing exclusively on the dipole-dipole interaction.
It appears that molecules in the highest weak-field-seeking 
states are mostly insensitive to short-range effects.

The ``raw'' scattering channels have the form
$|F_1M_{F_1}\epsilon_1\rangle|F_2M_{F_2}\epsilon_2\rangle|lm_l\rangle$,
which specifies the internal state of each molecule and the partial wave
$l$ describing the  relative orbital angular momentum of the molecules.
In a field, of course, the hyperfine and parity quantum numbers are no 
longer good.
It is therefore essential to consider a set of scattering
channels ``dressed'' by the appropriate field. This is achieved by 
diagonalizing the Stark or Zeeman Hamiltonian of each molecule,
including the $\Lambda$ doubling and hyperfine structure.  The resulting
eigenvectors then comprise the molecular basis used to construct
the scattering  Hamiltonian.  Field dressing is essential because otherwise
non-physical couplings between channels presist to infinite separation.  
The diagonal contributions of the Stark and Zeeman Hamiltonian in the 
field dressed basis define to the scattering thresholds as 
$R\rightarrow \infty$.

The channels involved in a given scattering process are further 
constrained by symmetries.  Namely, scattering of identical
bosons restricts the basis set to even values of $l$ only.
In addition, the cylindrical symmetry enforced by the external
field guarantees that the total projection of angular momentum
on the field axis, $M=M_{F_1}+M_{F_2}+m_l$, is a conserved quantity.

The scattering wave function is expanded in these field-dressed
channels, leading to a set of coupled-channel Schr\"{o}dinger
equations
\begin{equation}
\left({-\hbar^2\over2m_r}{d^2\over dR^2}\underline{1}+{\hbar^2 \hat l^2
\over2m_rR^2}\underline{1}+\underline{V}({\bf R})+\underline{H}_{FS}
\right)
\vec\psi(R)=E\vec\psi(R),\label{SE}
\end{equation}
where $\vec\psi$ is the multichannel wavefunction and $m_r$ is the reduced
mass.  The operators $\underline{H}_{FS}$ denotes the fine strucutre, 
including the effect of an electric or magnetic field.
(In this paper we do not yet include the simultaneous effect of both fields.)

In Eqn. (\ref{SE}), the operator $\underline{V}$ represents
the interaction between the molecules.  We are most interested in the
dominate dipole-dipole interaction whose general operator form is
\begin{equation}
 H_{\mu\mu}=-\
{3 (\hat {\bf R} \cdot  \hat{\bf \mu}_1)(\hat {\bf R} \cdot \hat
{\bf \mu}_2)-\hat{\bf \mu}_1 \cdot\hat{\bf\mu}_2 \over R^3}.
\label{fulldidi}
\end{equation}
where ${\bf \hat \mu}_i$ the electric dipole of molecule $i$, $R$ is the
intermolecular separation, and ${\bf\hat R}$ is the unit vector defining
the intermolecular axis.  This interaction is conveniently 
re-written in terms of tensorial operators as follows \cite{Brink}:
\begin{equation}
\label{didi}
 H_{\mu\mu}= -{\sqrt{6}\over R^3} \sum_{q} (-1)^q
C^2_q \cdot (\mu_1 \otimes  \mu_2)^2_{-q}.
\end{equation}
Here $C^2_q(\theta,\phi)$ is a reduced spherical harmonic that
acts only on the relative angular coordinate of the molecules,
while $(\mu_1 \otimes  \mu_2)^2_{-q}$ is the second rank
tensor formed from two rank-1 operators, $\mu_i$ that
act on the state of the $i^{th}$ molecule.  
These first rank operators are written as reduced spherical harmonics, 
$C_q^1(\alpha\beta)$ where $\alpha$ and $\beta$ are two of the Euler angles of
the rigid rotator wavefunction. With this form of the dipole-dipole 
interaction, we can then evaluate the matrix element. 

In the hyperfine parity basis (\ref{Fparitybasis}) the matrix 
elements are \cite{AA_PRA}
\begin{widetext}
\begin{eqnarray}
\langle 12lm_l|H_{\mu\mu}|1^\prime2^\prime l^\prime
m_l^\prime\rangle=&&{\sqrt{6}\mu^2_{{\emph E}}\over R^3}
[l,l^\prime,J_1,J^\prime_1,J_2,J^\prime_2,F_1,F^\prime_1,F_2,F^\prime_2]
\left({1+\epsilon_1\epsilon^\prime_1(-)^{J_1+J_1^\prime+2\Bar\Omega_1+1}
\over2}\right)
\left({1+\epsilon_2\epsilon^\prime_2(-)^{J_2+J_2^\prime+2\Bar\Omega_2+1}
\over2}\right)
\nonumber\\\times 
&&(-1)^{(1+F_1+F_1^\prime+F_2+F_2^\prime+
J_1+J_1^\prime+J_2+J_2^\prime+M_1+M_2-\Omega_1^\prime-\Omega_2^\prime+M_l)}
\nonumber\\\times
&&\left(\begin{array}
{ccc}1&1&2\\
M_{F_1}-M_{F_1^\prime}&
M_{F_2}-M_{F_2^\prime}&
M_{l}-M_{l}^\prime
\end{array}\right)
\left(\begin{array}{ccc}
J_1&1&J_1^\prime\\
-\bar\Omega_1&0&\bar\Omega_1
\end{array}\right)
\left(\begin{array}{ccc}
J_2&1&J_2^\prime\\
-\bar\Omega_2&0&\bar\Omega_2
\end{array}\right)
 \nonumber\\\times
&&\left(\begin{array}{ccc}
1&F_1&F_1^\prime\\
M_{F_1}-M_{F_1^\prime}&-M_{F_1}&M_{F_1^\prime}
\end{array}\right)
\left(\begin{array}{ccc}
1&F_2&F_2^\prime\\
M_{F_2}-M_{F_2^\prime}&-M_{F_2}&M_{F_2^\prime}
\end{array}\right)
\nonumber\\\times
&&\left(\begin{array}{ccc}
l^\prime&2&l\\
M_{l^\prime}&M_l-M_{l^\prime}&-M_l
\end{array}\right)
\left(\begin{array}{ccc}
l^\prime&2&l\\
0&0&0
\end{array}\right)
\left\{\begin{array}{ccc}
F_1&F_1^\prime&1\\
J_1^\prime&J_1&I
\end{array}\right\}
\left\{\begin{array}{ccc}
F_2&F_2^\prime&1\\
J_2^\prime&J_2&I
\end{array}\right\}.
\label{dipolehf}
\end{eqnarray}
\end{widetext}
A central feature of this matrix element is the factor
$\left(1-\epsilon_1\epsilon^\prime_1\right)\left(1-\epsilon_2\epsilon^\prime_2
\right)$, using $J_i=J_i^\prime=\bar\Omega_i=3/2$.  As a consequence of this 
factor, matrix elements diagonal
in parity identically vanish in zero electric field.  Instead,
for example, two $f$-parity states only interact with one another
via coupling to a channel consisting of two $e$-parity states. 

This dependence on parity is perhaps not unexpected, since 
the dipole-dipole force is of course transmitted by the dipole
moment of the first molecule producing an electric field
that acts on the second molecule.  But in a state of good
parity, the first molecule does not have a dipole moment until
it is acted upon by the second molecule.  Thus, both molecules
must simultaneously mix states of opposite parity to interact.  
Notice that in the
presence of an electric field, the dipoles are already partially polarized,
and this restriction need not apply; the scattering channels
are already directly coupled.  This change is of decisive importance
in elucidating the influence of electric fields on collisions.
In a magnetic field, by contrast, parity remains conserved and the
interactions are intrinsically weaker as a result.


\section{Inelastic Rates of OH-OH collisions in external fields}

We move now to the consequences of the interaction (\ref{dipolehf})
on scattering.  Scattering calculations are done using the 
log-derivative propagator method \cite{johnson}.  To ensure
convergence at all collision energies and applied fields, it was
necessary to include partial waves up to $l=6$, and to carry
the propagation out to an intermolecular distance of $R=10^4 (a.u.)$
before matching to long-range wave functions.  Cross sections and
rate constants are computed in the standard way for anisotropic 
potentials \cite{Bohn00PRA}.

We remind the reader that throughout we consider collisions
of molecules initially in their $|FM_F\epsilon \rangle =$
$|22 - \rangle$ states, which are weak-field seeking for both
electric and magnetic fields.  Thus for a scattering
process incident on an s- partial wave, the incident
channel will be written $|i \rangle = |F_1M_{F_1}\epsilon_1 \rangle
|F_2M_{F_2}\epsilon_2\rangle |lm_l \rangle =$
$|22,-\rangle |22,- \rangle |00 \rangle$. 

In the following, we will make frequent reference to ``energy
gap suppression'' of collision rates.  This notion arises from
a perturbative view of inelastic collisions, in which case
the transition probability amplitude is proportional to the
overlap integral
\begin{equation}
\label{overlap}
\int dR \psi_i(R) V_{if}(R) \psi_f(R)
\end{equation}
where $\psi_{i,f}$ denote the incident and final channel radial
wave functions, and ${V}_{if}$ is the coupling matrix element
between them.  In our case, $\psi_i$ will have a long de Broglie
wavelength corresponding to its essentially zero collision
energy. The de Broglie wavelength of $\psi_f$ will instead 
grow smaller as the energy gap $E_i - E_f$ between incident
and final thresholds grows.  Thus the integral in (\ref{overlap}),
and correspondingly the collision rates, will diminish.
For this reason, the collisions we consider tend to favor
changing the hyperfine states of the molecules over changing
the parity states, since the hyperfine splitting of OH
is smaller than the $\Lambda$-doubling.

\subsection{Electric Field Case}
\begin{figure}
\centerline{\epsfxsize=5.0cm\epsfysize=5.0cm\epsfbox{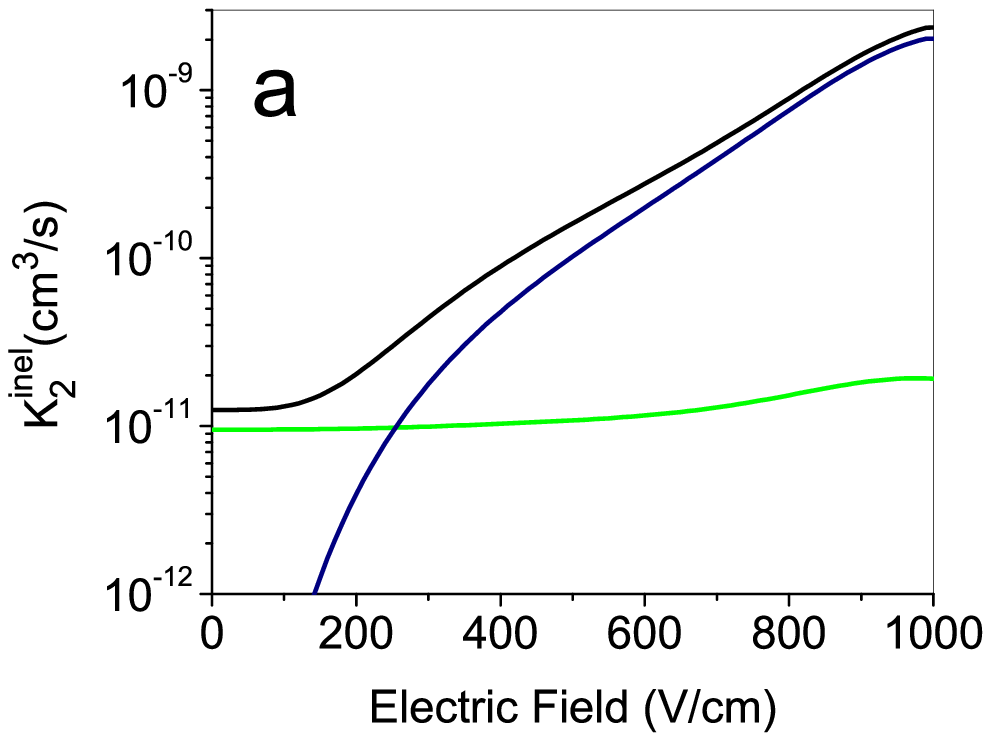}}
\centerline{\epsfxsize=5.0cm\epsfysize=5.0cm\epsfbox{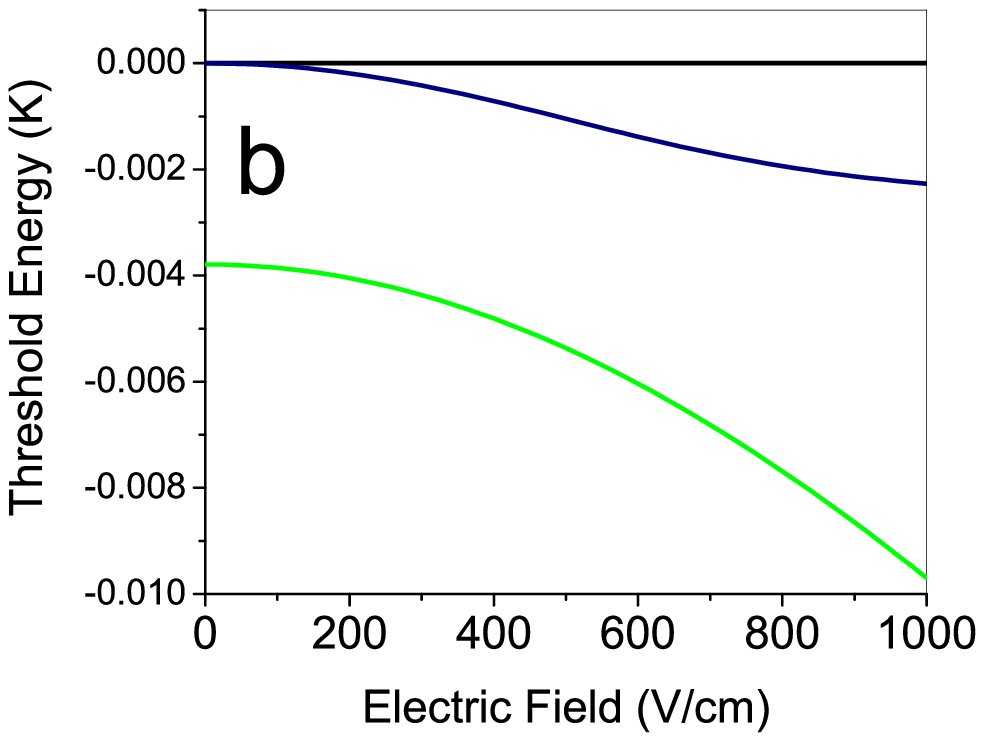}}
\caption{(a) Total (black) and selected partial  (color) inelastic 
rates for
OH-OH collisions as a function of electric field. The green 
curve is the dominant zero field inelastic loss process to channel
$|10-\rangle|22-\rangle$ (green).  In the presence of the field, a different 
channel, $|21-\rangle|22-\rangle$, becomes dominant (blue).
(b) The thresholds for these exit channels, relative to the 
incident threshold.
} \label{eprates}
\end{figure}

To calculate scattering in the presence of an electric field, we 
only need to include partial waves 
$l=0,2$ for numerical accuracy of $K^{inel}_2$ for the field range that 
we consider, $E\leq 1000 (V/cm)$, and at a collision energy of
$10^{-5}$ K.  Here we are only interested 
in the trend and identification of the loss mechanism.  To numerically 
converge the inelastic rates at higher field
values, where the induced dipoles are large, naturally requires 
more partial waves.

Figure \ref{eprates} (a) shows the total (black)
and partial  (color) inelastic rate constant $K_2^{inel}$ as a function of 
the electric field (compare Fig. \ref{temp} (a)).  
Even in zero field, where the dipolar forces nominally average out, 
the rate constant is large, comparable to the elastic rate
constant.  This fact attests to the strength of dipolar
forces in OH, even in second order.  

The green line in Fig. \ref{eprates}(a) represents losses to the dominant 
zero field loss channel $|10-\rangle|22-\rangle|22\rangle$.  
The blue curve in Fig. \ref{eprates} (a) represents instead
the dominant loss process at higher electric field values,
in channel $|21- \rangle |22- \rangle |21 \rangle$.  Whereas the
former rate remains relatively insensitive to field, the latter
rises dramatically.

This behavior arises from two competing tendencies in an
electric field.  The first is the increasing mixing of different
parity states as the field is turned on, leading to an
increasing strength of the direct dipole-dipole coupling
that affects both exit channels.
This additional coupling would, in general, cause inelastic
rates to rise.  It is, however, offset by the competing tendency
for inelastic rates to become less likely when the change
in relative kinetic energy of the collision partners is larger.
Fig. \ref{eprates}(b) shows the threshold energies for the two
exit channels in Fig. \ref{eprates}(a), versus field,
with zero representing the energy of the incident threshold.  Here it
is evident that loss to the channel $|22-\rangle |10-\rangle|22\rangle$
(green line) is accompanied by a large gain in kinetic energy,
whereas loss to channel $|21- \rangle |22- \rangle |21 \rangle$
(blue line) gains comparatively little kinetic energy, and
thus the later channel more strongly affected by the increased 
coupling generated by the field.

\subsection{Magnetic Field Case}
\begin{figure}
\centerline{\epsfxsize=5.0cm\epsfysize=5.0cm\epsfbox{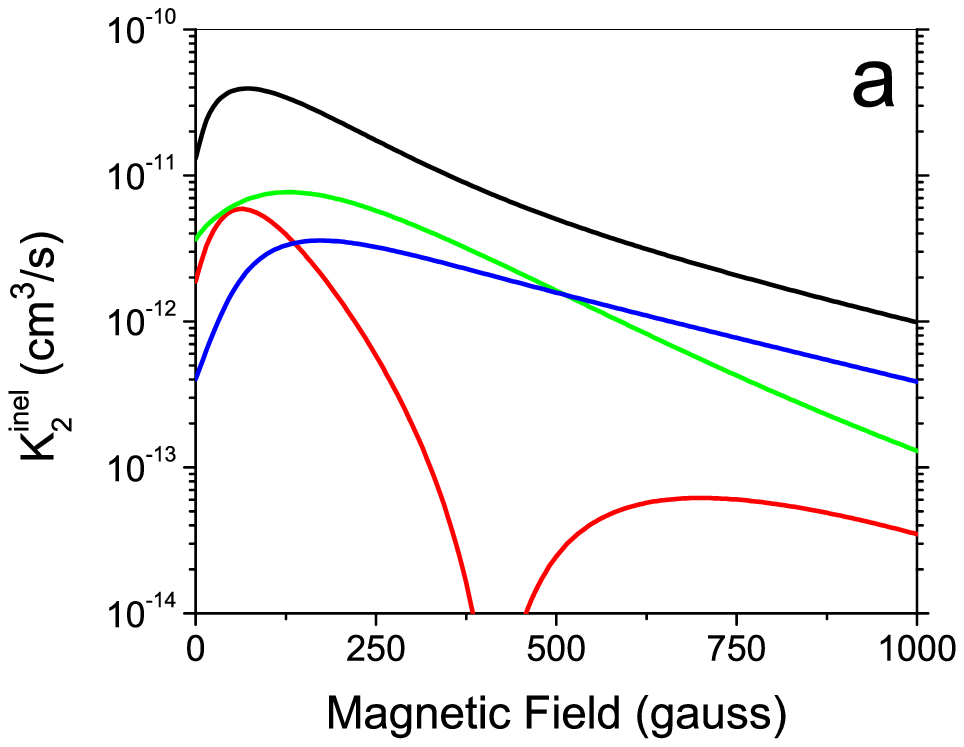}}
\centerline{\epsfxsize=5.0cm\epsfysize=5.0cm\epsfbox{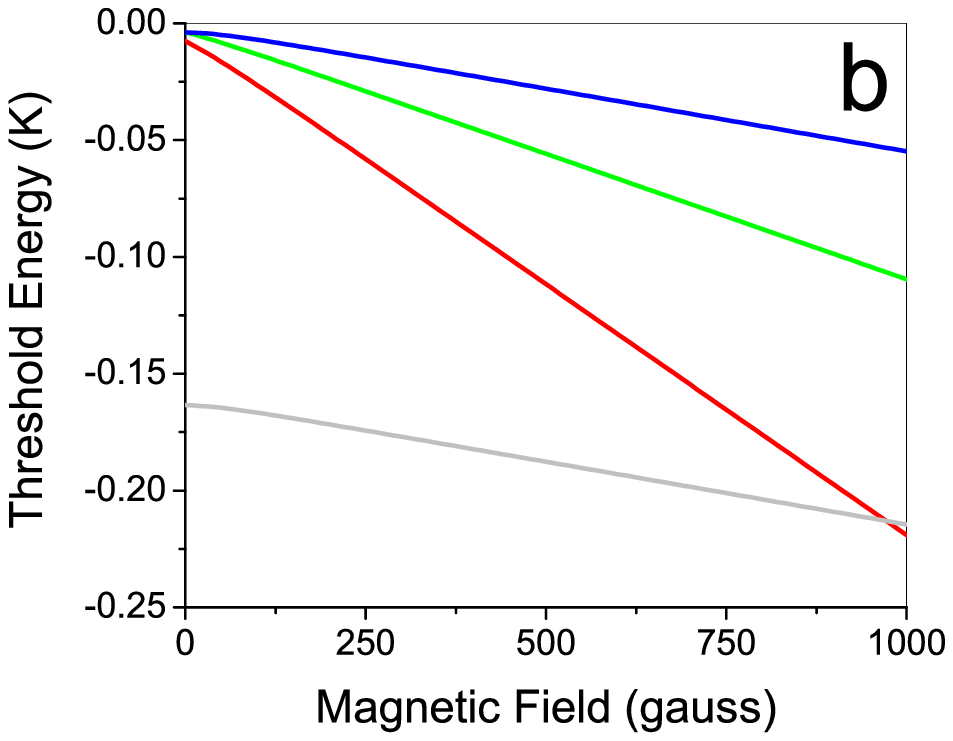}}
\caption{(a) The total (black) and partial (color) inelastic rates 
for OH-OH collisions as a function of magnetic field.   
The colors are explained in the text.
(b) The  corresponding thresholds, referred to the 
incident channel's threshold ($E_i=0$).  The lowest curve is one 
possible intermediate channel.} \label{prates}
\end{figure}

To gain insight into the suppression of the inelastic rates in a magnetic
field (Fig. \ref{prates} (a)), 
calculations were at a representative collision energy  
$E=10^{-5} K$.  To converge
the calculations in high field (B$\geq1500$ gauss) required partial
waves $l=0,2,4,6$.  We have only considered collisions with incident
partial wave $l=0$, since higher partial wave contributions, while
they exist, only contribute to rates at the fraction of a precent level.

Because the electric field remains zero,  parity is still
a rigorously good quantum number.  Therefore states of the
same parity are not directly coupled.  Nevertheless, the dominant
loss channels in a magnetic field share the parity of the incident
channel wave function, 
$|i \rangle = |22,-\rangle |22,- \rangle |00 \rangle$. 
Figure \ref{prates} (a) illustrates this by showing the total (black)
and partial (color) inelastic rates as a function of the magnetic field.
The loss rates shown correspond to the exit channels $|f \rangle =$
$|10-\rangle|22-\rangle|22\rangle$(green),
$|11-\rangle|22-\rangle|41\rangle$(blue), and 
$|10-\rangle|10-\rangle|44\rangle$(red).

Since direct coupling to these final channels is forbidden to the
dipolar interaction, all coupling must occur through some
intermediate channel $|int \rangle$.  Moreover, owing to the
parity selection rules in the matrix element (\ref{dipolehf}),
this intermediate channel must have parity quantum numbers 
$\epsilon_1 = \epsilon_2 = +$.  
Since this coupling is second order, the
dominant exit channels can consist of both d-wave ($l_f=2$) and 
g-wave ($l_f=4$)
contributions, in contrast to the electric field case.

The primary feature of the inelastic rates in Fig. \ref{prates} (a)
is that they decrease significantly 
at large field. This decline is the main reason for optimism regarding 
evaporative cooling strategies in OH; an applied bias field of
3000 Gauss can reduce the inelastic rate constant to below 
$2\times10^{-13}$ cm$^3$/sec, see Figure \ref{temp}.
The cause of this decrease can be traced directly to the relative 
separation of the incident and final channel thresholds, along with
the indirect nature of the coupling.  

To see this, we reduce the model to its essential ingredients:
(1) a strong dipole-dipole
interaction, (2) the relative motion of the thresholds as the
magnetic field is tuned, (3) an extremely exothermic 
intermediate channel and (4) the centrifugal barrier in the final
and intermediate channels.  The Hamiltonian for a reduced model is 
$H_{Model}=T_0+V_{Model}$, where $T_0$ is the kinetic energy operator 
and $V_{Model}$ in matrix form is
\begin{eqnarray}
V_{Model}=\left(
\begin{array}{ccc}
E_{i}&0& a/R^3 \\
0&E_{f}+c_f/R^2&b/R^3\\
a/R^3&b/R^3&E_{int}+c_{int}/R^2\\
\end{array}\right). \label{model}
\end{eqnarray}
Here $c_j$ is a centrifugal repulsion $c_j=\hbar^2 l_j(l_j+1)/2m_r$, $a$ and 
$b$ are  dipole-dipole coupling strengths, and $E_j$ are the threshold
energies for the $j^{th}$ channel.  
The channels $\{i,f,int\}$ have quantum numbers
$(\epsilon_1\epsilon_2)^i=(\epsilon_1\epsilon_2)^f=(--)$ and 
$(\epsilon_1\epsilon_2)^{int}=(++)$.  The incident channel has
partial wave $l_i=0$, while dipole coupling selection rules
allow $l_{int}=2$, and $l_f=2$ or $l_f=4$.

The model Hamiltonian (\ref{model}) explicitly excludes direct coupling
between incident and final channels, whereas coupling is mediated
through the $int$ channel. 
Parameters characteristic of the physical problem are
$a=0.12$ (a.u.),  $b=0.10$ (a.u.),  
$E_i=0$, $E_f=-0.003\rightarrow-0.1(K)$, and
$E_{int}=-0.17(K)$,
$l_i=0$, $l_f=2$ or $l_f=4$ and $l_{int}=2$.  Because of the
energy gap separation losses to the
intermediate channels are negligible.  We find, in addition,
that moving $E_{int}$ has little effect on the rate constants
for loss to channel $f$.

\begin{figure}
\centerline{\epsfxsize=5.0cm\epsfysize=5.0cm\epsfbox{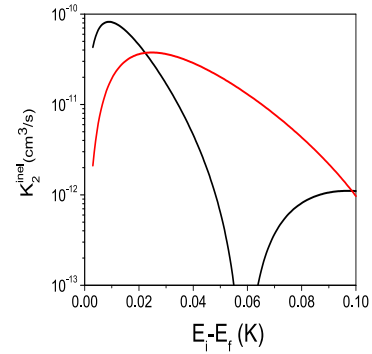}}
\caption{ Inelastic rate constants for the 3-channel model system,
Eq. \ref{model}, as function of initial and final threshold separation.
The two curves are for $d$ and $g$-wave exit channels black and red,
respectively. In $g$-wave channel $K_2^{inel}$ evolves
more slowly as the thresholds are separated.} 
\label{modelfig}
\end{figure}

Fig. \ref{modelfig} shows the inelastic rates computed within this model.
This three channel model does a reasonable job of mimicking the prominent 
features of the full calculation, including 
the eventual and lasting decrease in rates as the states are separated
in energy. In addition, the g-wave rates decay more slowly as
a function of field than do the d-wave rates, consistent with the
full calculation (compare Fig. \ref{prates}).  The declining values of 
the rate constant cannot, however, be attributed to a simple
overlap integral of the form (\ref{overlap}), since the incident
and final channels are not directly coupled.  We therefore 
present a more refined adiabatic analysis of this process
in the next subsection.

\subsection{Adiabatic Analysis of the magnetic field case}

To understand the system's magnetic field behavior 
we analyze the reduced channel model (\ref{model}) 
in the adiabatic representation \cite{AA_FL,child}.
This representation assumes that $R$ is a ``slow'' coordinate. 
At every $R$ we diagonalize the Hamiltonian in all remaining 
degrees of freedom.  Since it is not rigorously true that $R$
varies infinitely slowly, the residual nonadiabatic couplings 
can be accounted for in the kinetic energy operator.  
Written more 
formally we diagonalize
\begin{equation}
\underline{W}=\left({\hbar^2l(l+1)
\over2m_rR^2}\underline{1}+\underline{V}(R)+\underline{H}_Z\right) 
\label{ad}
\end{equation}
where the terms are the centrifugal barrier, potential matrix including 
dipole-dipole interaction and the Zeeman Hamiltonian.
Diagonalizing the matrix in Eq. \ref{ad}, we get
$\underline{W}|\alpha(R)\rangle=U_\alpha(R)|\alpha(R)\rangle$ 
where $U_\alpha(R)$ are the eigenvalues and 
$|\alpha(R)\rangle$ are the eigenvectors. With the eigenvectors 
we are able to form a linear transformation $\underline{X(R)}$ which transforms
between the diabatic and adiabatic representations, ie
$\underline{X^T}\underline{W}(R)\underline{X}=\underline{U}(R)$.  The 
eigenvalues and eigenvectors have radial dependence, but for notational 
simplicity $(R)$ will be suppressed hereafter.

To distinguish between adiabatic and diabatic representations we use 
Greek letters 
($\alpha, \beta, ...$) to denote the 
adiabatic channels and Roman letters (i,j,...) to denote diabatic channels.
When considering specific inelastic processes in the diabatic basis we denote 
initial and final channels as $i$ and $f$ and for the adiabatic channels
as $\iota$ and $\varphi$.  In the limit $R \rightarrow \infty$,
the two sets of channels coincide.

\begin{figure}
\centerline{\epsfxsize=5.0cm\epsfysize=5.0cm\epsfbox{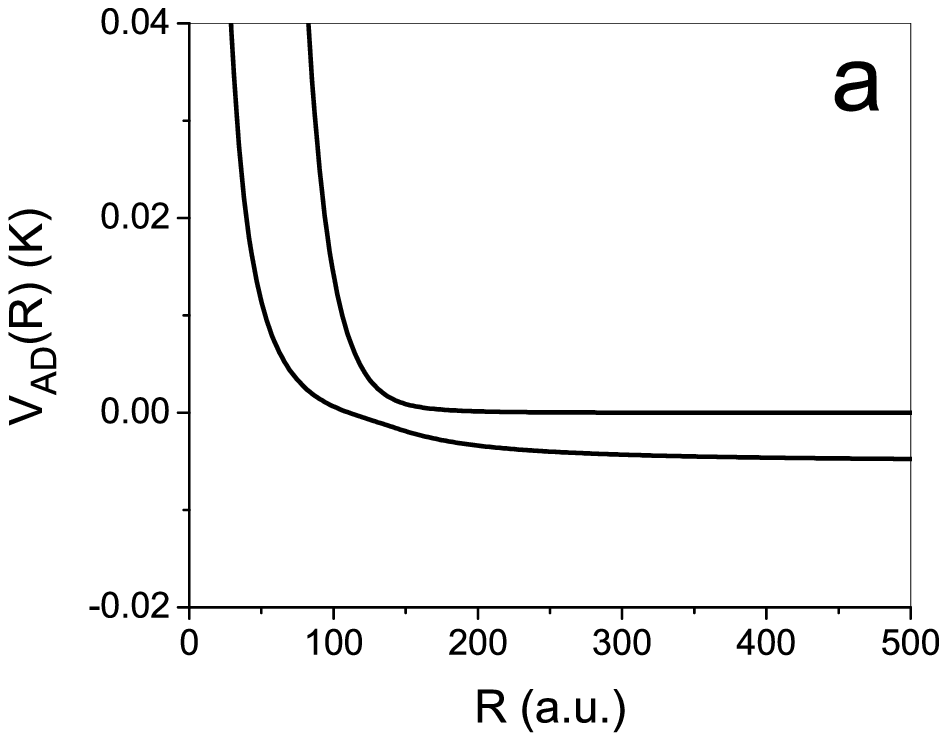}}
\centerline{\epsfxsize=5.0cm\epsfysize=5.0cm\epsfbox{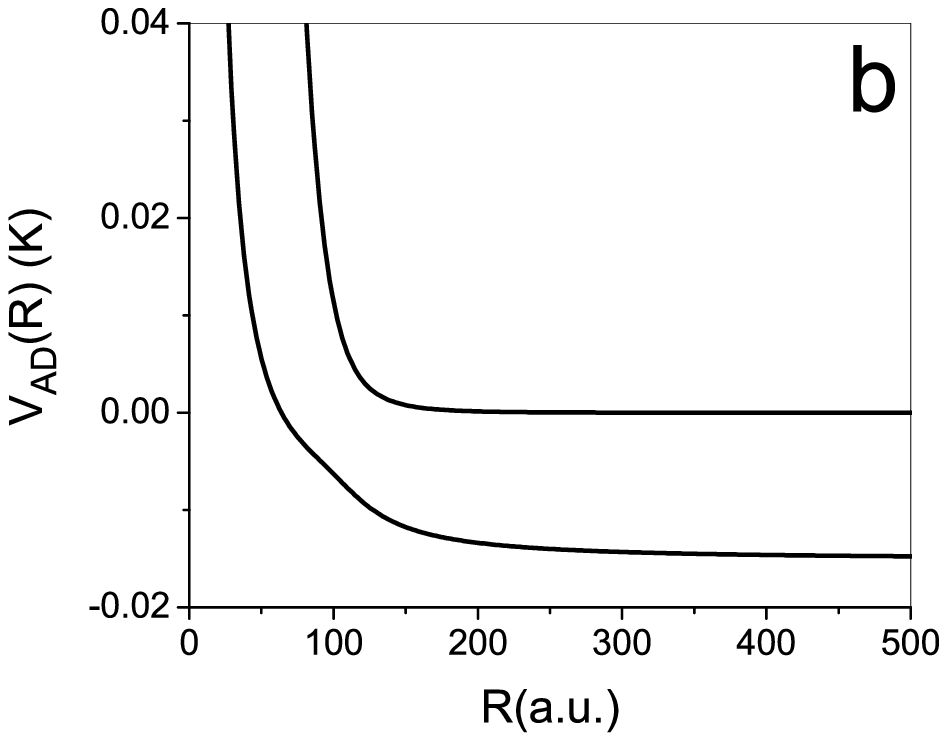}}
\caption{The relevant adiabatic potential curves for the 
OH-OH system.  Shown are two different values of the final
threshold energy $E_i-E_f=5mK$ (a)  and $E_i-E_f=15mK$ (b)}
\label{curves}
\end{figure}

A partial set of adiabatic potential curves generated in this way is shown
in Fig. \ref{curves}, exhibiting an avoided crossing at $R = 150$.
Thus molecules incident on the
uppermost channel scatter primarily at large values of $R$.  This
point has been made in the past when an electric field is applied 
\cite{AA_PRA}; here we note that it is still true in zero electric field,
and that scattering calculations can proceed without reference to
short-range dynamics.

The transformation between the representations is $R$ dependent  
implying that the channel couplings shift from the potential to the 
kinetic 
energy operator.  Using the adiabatic representation changes Eq. \ref{SE} to 
\begin{eqnarray}
\left(\underline{X}^T{-\hbar^2\over2m_r}{d^2\over
dR^2}\underline{X}+\underline{U}\right)\vec\xi(R)=
\nonumber\\ \left({-\hbar^2\over2m_r}
\left({d^2\over dR^2}+2\underline{P}{d\over dR}+\underline{Q}\right)
+\underline{U}\right)
\vec\xi(R)=E\vec\xi(R).
\end{eqnarray}
Here $\vec\xi=\underline{X}^T\vec\psi$. 
 
To get the channel couplings in the adiabatic picture we need
matrix elements of the derivative operators, defined as
$P_{\alpha\beta}=\langle \alpha|{d\over dR}| \beta\rangle$. 
We evaluate the the $P_{\alpha\beta}$ matrix, the
dominant off-diagonal channel coupling, using the Hellmann-Feynman
theorem \cite{child}
\begin{equation}
P_{\alpha\beta}={\sum_{kl}X_{\alpha k}^T 
\nabla V_{kl} X_{l\beta}\over U_\alpha-U_\beta}.  
\end{equation}
\begin{figure}
\centerline{\epsfxsize=5.0cm\epsfysize=5.0cm\epsfbox{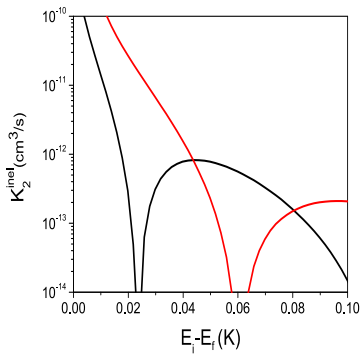}}
\caption{Inelastic rate constants as estimated by the adiabatic 
distorted wave Born approximation for the 3-channel system. 
The black curve is for a d-wave exit channel and the red for a 
g-wave exit channel.}
\label{adborn}
\end{figure}

Scattering amplitudes are then easily estimated in the adiabatic
distorted-wave Born approximation (ADWBA).  Namely, we construct
incident and final radial wave functions $\phi_{\iota,\varphi}$
that propagate according to the adiabatic potentials $U_{\iota,\varphi}$.
In terms of these adiabatic wavefunctions, the scattering $T$-matrix
is given by an overlap integral analogous to Eqn. \ref{overlap}
\begin{equation}  
T_{\iota\varphi}={\pi\hbar^2\over m_r}\langle
\phi_{\iota}|{\overleftarrow{d/dR}\underline{P}
+\underline{P}\overrightarrow{d/dR}\over
\sqrt{2}} |\phi_{\varphi}\rangle.
\end{equation}  
Here $\overleftarrow{d/dR}$
($\overrightarrow{d/dR}$) is the radial derivative operator acting
to the left (right).  The cross section for identical bosons is
$\sigma_{\iota \varphi}={8\pi\over k_\iota^2}|T_{\iota\varphi}|^2$.  
From here we are able to numerically 
calculate a rate constant for inelastic loss $K_2^{inel}=v_\iota
\sigma_{\iota\varphi}$ where $v_\iota$ is the asymptotic velocity 
given by $\sqrt{2U_\iota(R\rightarrow \infty)/m_r}$. 

The result of the ADWBA is shown in figure \ref{adborn}. The two curves are
for $d$ (black) and $g$-wave (red) inelastic channels.  Several key features
are present that also occur in the full calculation, namely:
(1) the inelastic rate goes down with increasing threshold separation.
(2) there is a zero in the rates as seen in Fig. \ref{prates}
(3) the $g$-wave inelastic rate goes more slowly than the $d$-wave as seen
in the model and the full calculation.  The ADWBA accounts for all of these.
The first feature, diminishing rates, still arises from an energy
gap suppression, since the de Broglie wavelengths of incident and
final channel still do not match well.  In the ADWBA this process is further
helped along by the fact that the residual channel coupling,
represented by $P$, is localized near the avoided crossings 
of the adiabatic potential curves.

\begin{figure}

\centerline{\epsfxsize=5.0cm\epsfysize=5.0cm\epsfbox{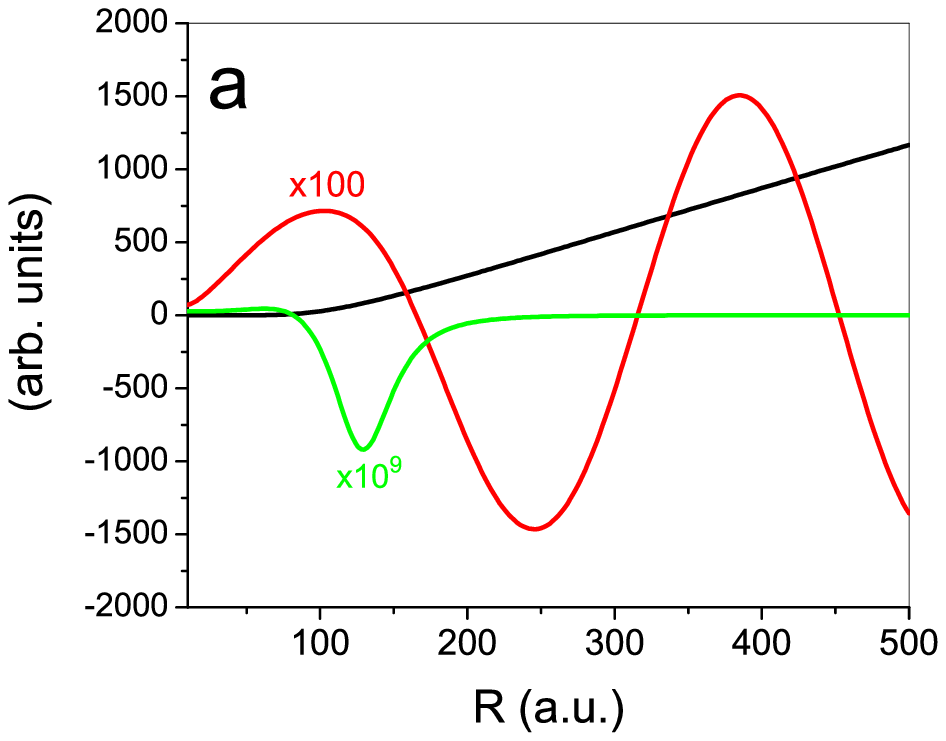}}
\centerline{\epsfxsize=5.0cm\epsfysize=5.0cm\epsfbox{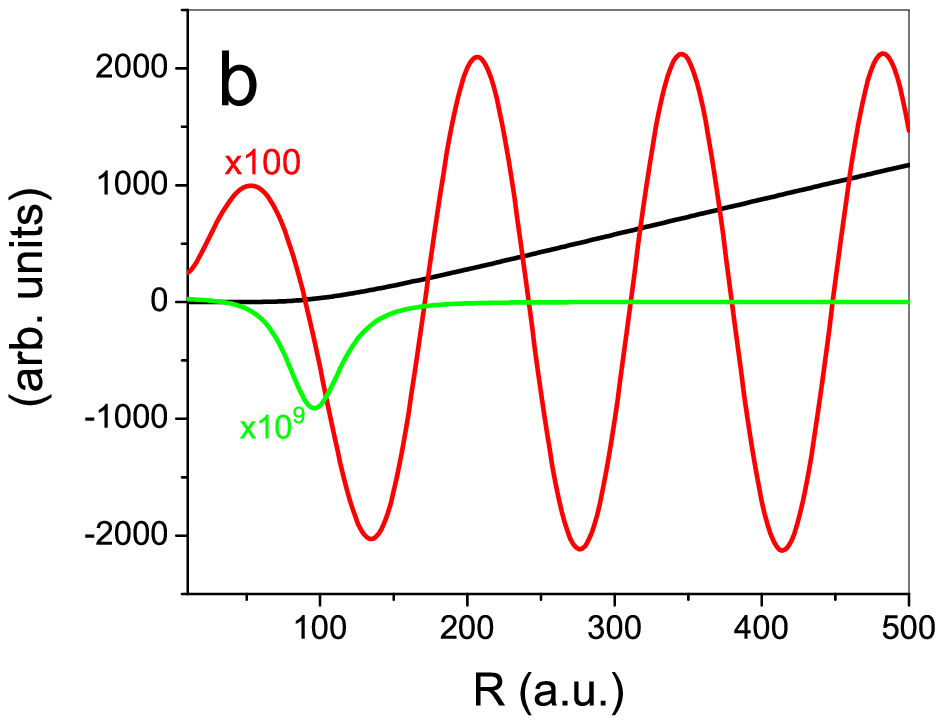}}
\centerline{\epsfxsize=5.0cm\epsfysize=5.0cm\epsfbox{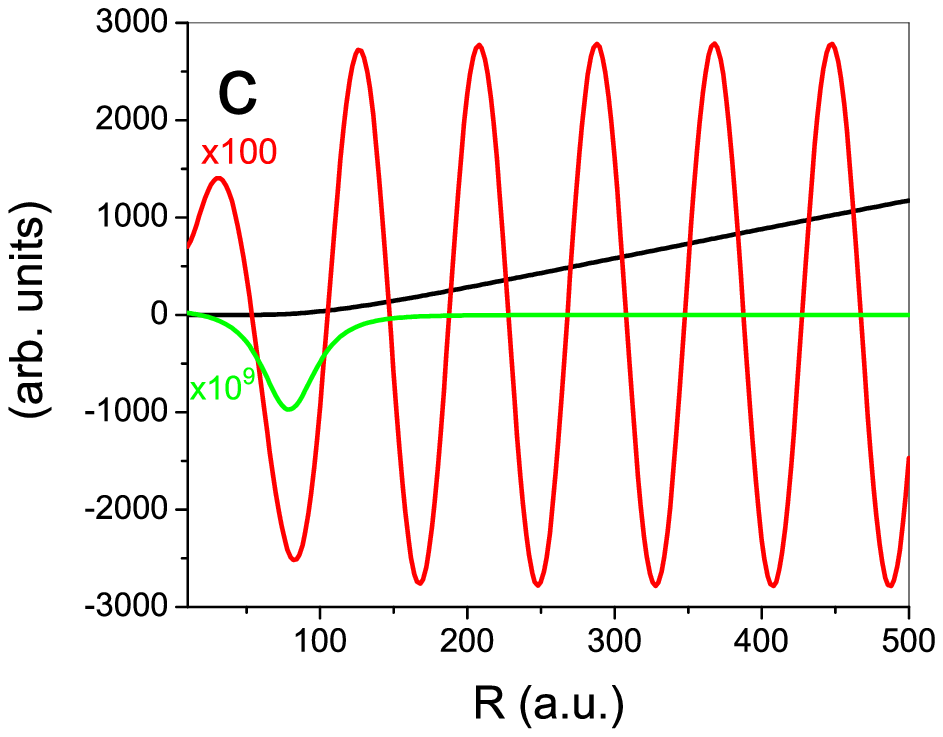}}
\caption{Illustrations of the origin of the zero in the partial 
rates.  Each panel shows curves 
$\psi_\iota$, ${d \over d R}\psi_\varphi\times100$ and  $P_{\iota\varphi}$ 
$\times10^{9}$. The plots are for different final energies and inelastic
rate from the Born approximation are
(a) $E_f=-6mK$ and $K^{inel}_{\iota\varphi}=5\times10^{-11}$ 
(b) $E_f=-22mK$ , $K^{inel}_{\iota\varphi}=5\times10^{-14}$
(c) $E_f=-62mK$, $K^{inel}_{\iota\varphi}=6\times10^{-13}$,
see text for details. }
\label{wave}
\end{figure}

The ADWBA helps to visualize this suppression, as shown by the sample
wave functions in Fig. \ref{wave}.  This figure 
shows $\psi_\iota$, ${d\over dR}\psi_\varphi$, and 
$P_{\iota\varphi}$ for various values of $E_f$.  Varying $E_f$ mimics  
the threshold motion of the system.  The values of $E_f$
of Figure \ref{wave} are
(a) $E_f=-6mK$,  (b) $E_f=-22mK$, and (c) $E_f=-62mK$.  The effect of
the different $E_f$s leave $\psi_\iota$ mostly unchanged, however
$\psi_\varphi$ becomes
more exothermic and therefore more oscillatory 
($\lambda_{db}$ clearly shortens).  Moreover,
the dominant coupling region, where $P_{\iota\varphi}$ peaks, 
moves to shorter $R$ as $E_f$ increases.  This motion is obvious
from the avoided crossing in Fig. \ref{curves}.

The transition amplitude in the ADWBA is proportional to the integral
of the product of the three quantities in Fig. \ref{wave}.
Because of the shortening of the de Broglie wavelength in the exit
channel, this integral will eventually vanish, accounting for the
zero in the inelastic rates.  The {\it total} rate will, in general,
not vanish, since there are many exit channels, and they
will experience the destructive interference at different values
of the threshold, hence at different fields.

Finally, the $g$-wave inelastic rates are not so strongly affected by 
the separation of $E_i$ and $E_f$
because the g-wave centrifugal barrier is larger, meaning a greater
energy is required to change the wave function at short range such that 
velocity node can pass through the coupling region.  The zero
in this rate constant will thus occur at larger threshold separations.

\section{Conclusions }

We have explored the influence of a magnetic field on the cold
collision dynamics of polar molecules.  The dipole-dipole 
interactions remain significant even in the absence of an
electric field that polarizes the molecules.  In general
this implies that molecular orientations are unstable in
collisions, making magnetic trapping infeasible.  We have found,
however, that a suitably strong magnetic field can mitigate
this instability.  

Beyond this result, we note that laboratory strength fields can
exert comparable influence on cold collisions, if applied separately.
A useful rule of thumb in this regard is that an electric field of 
300 $V/cm$ acting on a 1 $Db$ dipole moment caues roughly the same 
energy shift as a 100 $gauss$ field acting on a 1 Bohr magneton magnetic
moment.
This raises the interesting question of how the two fields
can be applied simultaneously, to exert even finer control over
collision dynamics.  This will be the subject of future investigations.

\begin{acknowledgments}
This work was supported by the NSF and by a grant from the 
W. M. Keck Foundation.
\end{acknowledgments}
\bibliographystyle{amsplain}

\end{document}